\documentclass[lettersize,journal]{IEEEtran}
\usepackage{amsmath,amsfonts}
\usepackage{algorithmic}
\usepackage{array}
\usepackage[caption=false,font=normalsize,labelfont=sf,textfont=sf]{subfig}
\usepackage{textcomp}
\usepackage{stfloats}
\usepackage{url}
\usepackage{verbatim}
\usepackage{graphicx}
\usepackage{balance}
\usepackage{xcolor}
\usepackage[normalem]{ulem}

\newcommand{\mat}[1]{\mathbf{#1}}

\def\BibTeX{{\rm B\kern-.05em{\sc i\kern-.025em b}\kern-.08em
    T\kern-.1667em\lower.7ex\hbox{E}\kern-.125emX}}

\begin{document}
\title{p-DCIM: A 28nm Probabilistic SRAM DCIM based Ising Machine for Near-Memory Context Selection in LLM-RAG Pipeline}
\author{\IEEEauthorblockN{
Madhav Vadlamani\textsuperscript{*, 1},
Rahul Singh\textsuperscript{*, 2},
Yuyao Kong\textsuperscript{1},
Haomei Liu\textsuperscript{1},
Zheng Zhang\textsuperscript{2},
Shimeng~Yu\textsuperscript{$\dagger$,1}
} \\
\IEEEauthorblockA{\textsuperscript{1}School of Electrical and Computer Engineering,
Georgia Institute of Technology, Atlanta, GA, 30332} \\
\IEEEauthorblockA{\textsuperscript{2}Department of Electrical and
Computer Engineering, University of California, Santa Barbara, CA, 93106} \\
\IEEEauthorblockA{\textsuperscript{*}Equal contribution.}
\IEEEauthorblockA{$^{\dagger}$Corresponding author: Shimeng Yu (shimeng.yu@ece.gatech.edu).}

\thanks{Manuscript created October, 2020; This work was developed by the IEEE Publication Technology Department. This work is distributed under the \LaTeX \ Project Public License (LPPL) ( http://www.latex-project.org/ ) version 1.3. A copy of the LPPL, version 1.3, is included in the base \LaTeX \ documentation of all distributions of \LaTeX \ released 2003/12/01 or later. The opinions expressed here are entirely that of the author. No warranty is expressed or implied. User assumes all risk.}}

\markboth{Journal of \LaTeX\ Class Files,~Vol.~18, No.~9, September~2020}%
{How to Use the IEEEtran \LaTeX \ Templates}

\maketitle

\begin{abstract}
We present a fabricated SRAM DCIM Ising machine that performs set-level
context selection for large language model (LLM) with retrieval-augmented generation
(RAG). In RAG, the evidence-selection stage is increasingly delegated
to repeated language-model reranking, which adds substantial energy to the
inference pipeline. We reformulate evidence selection as a quadratic
unconstrained binary optimization (QUBO) problem and solve it on a dual-mode
digital compute-in-memory (DCIM) macro: the same multiply-and-accumulate
datapath computes the embedding-similarity scores that populate the QUBO
coefficients and evaluates the per-spin Hamiltonian differences that drive the
anneal. The annealing temperature is set by a digital plane-count control along with the SRAM pseudo-read random bit flips, giving graded cooling entirely from
on-chip digital configuration. The macro is
taped-out in a 28\,nm process [$0.16$\,mm$^2$, $0.41$\,mW] and is proposed to be integrated as part of the logic base-die in hybrid high-bandwidth memory (HBM) and high-bandwidth flash (HBF) stacks in a GPU package, where only the selected passage identifiers
return to the host. Replacing the language-model selection pass with the near-memory
Ising solver yields a macro-normalized selection-energy estimate
more than six orders of magnitude lower than the LLM-based baseline
and reduces the evaluated total energy per query by $67\%$, while
maintaining comparable token-level F1 on the evaluated examples.
\end{abstract}


\section{Introduction}
\IEEEPARstart{L}{arge} language models (LLMs) have demonstrated strong capabilities in natural-language understanding and generation, but their responses remain limited by the information encoded in their parameters and by the context supplied during inference. Retrieval-augmented generation (RAG) addresses this limitation by grounding generation in information retrieved from an external knowledge source, enabling access to dynamically updated or domain-specific content without modifying the underlying model~\cite{NEURIPS2020_6b493230,10.1145/3644815.3644945, Gao2023RetrievalAugmentedGF}. A typical RAG pipeline first retrieves a candidate set of passages from a vector database, selects or reranks a smaller subset, and then supplies the resulting context to the LLM for answer generation. As model sizes, knowledge bases, and retrieved candidate pools continue to increase, however, the intermediate context-selection stage becomes an increasingly important determinant of both answer quality and system cost.

Conventional retrieval pipelines commonly select the top-$K$ passages according to independent relevance scores. Although computationally simple, this approach does not explicitly account for interactions among the selected passages. In multi-hop and evidence-intensive question answering, the desired context is inherently a set-level object: multiple passages may provide redundant information, while individually lower-ranked passages may contain complementary evidence required to satisfy different information needs of the query~\cite{yang-etal-2018-hotpotqa, lee-etal-2025-shifting}. Recent approaches therefore perform set-wise context selection using an LLM, allowing the selector to reason jointly over candidate passages and query requirements~\cite{lee-etal-2025-shifting}. Such methods can improve evidence construction, but they introduce another transformer-inference pass into the RAG pipeline. For large models, this additional stage requires repeated movement of model weights between high-bandwidth memory (HBM) and the GPU and consequently incurs substantial latency and energy even though its output is only a small subset of passage identifiers.

These characteristics motivate separating semantic processing from combinatorial context selection. Recent work has formulated requirement-aware RAG evidence selection as a quadratic unconstrained binary optimization (QUBO) problem, in which binary variables indicate selected passages and the objective jointly captures passage relevance, requirement coverage, support strength, redundancy, complementarity, and context-size constraints~\cite{singh2026qubooptimizedevidenceselectionretrievalaugmented}. Unlike independent top-$K$ ranking, the resulting objective explicitly represents pairwise interactions among candidate passages and therefore optimizes the utility of the selected set as a whole. Once written in QUBO form, the selection problem can be executed using a broad class of Ising machines. This creates an opportunity to replace a computationally expensive LLM-based selection pass with an Ising machine dedicated to binary energy minimization.

Ising machines address combinatorial optimization problems by evolving toward low-energy spin configurations, and prior Ising and p-bit solvers have reported substantial energy and execution-time benefits on classical benchmarks such as the traveling-salesman, MAX-CUT, and satisfiability problems~\cite{aadit2022massively,9062965,10.3389/fphy.2019.00048,9720514}. SRAM-based digital compute-in-memory (DCIM) implementations provide a CMOS-compatible realization of this paradigm by storing the coupling coefficients directly in memory and evaluating the weighted sums required for spin updates in situ, thereby reducing repeated movement of the QUBO matrix between memory and arithmetic units~\cite{10.1145/3649329.3655673,Kong2026Compact}. These architectures have been demonstrated primarily on conventional combinatorial-optimization workloads such as the traveling-salesman problem.

In our recent work, we extended the SRAM DCIM Ising architecture from problem-specific combinatorial optimization toward a more general QUBO-solving framework, enabling arbitrary QUBO coefficient matrices to be mapped onto the in-memory annealing datapath and demonstrating the architecture on binary neural network robustness verification~\cite{vadlamani2026scalabledigitalcomputeinmemoryising}. In that application, adversarial counterexample search was formulated as a QUBO and executed using the DCIM Ising machine, establishing the applicability of the architecture to structured AI-derived optimization problems beyond classical benchmarks. The annealing stochasticity is generated directly from voltage-controlled SRAM pseudo-read errors, eliminating the need for dedicated noise-generation peripherals such as linear-feedback shift registers (LFSRs) or other external random-number-generation circuitry. The present work advances this direction further by targeting a recurring optimization stage within an LLM inference pipeline and by incorporating not only QUBO solution, but also hardware-efficient construction of the QUBO itself. In contrast to our previous BNN-verification study, which relied on simulation and PPA projections, the present work is supported by a new fabricated 28-nm SRAM-DCIM chip, with measured stochastic characteristics, power, and on-chip QUBO-solving operation.

RAG context selection introduces a different hardware challenge. Constructing the selection objective itself requires query-to-passage and passage-to-passage similarity evaluations over the retrieved candidate embeddings. For $N$ candidates, the pairwise terms can require $\mathcal{O}(N^2)$ embedding interactions before annealing begins. If these embeddings are first retrieved from storage and transferred to the GPU merely to construct the QUBO, much of the data-movement cost that an Ising accelerator is intended to avoid is incurred before optimization starts. This issue becomes particularly relevant in emerging near-storage RAG architectures such as HAVEN~\cite{hsu2026havenhighbandwidthflashaugmented}, where vector retrieval is performed directly within a High-Bandwidth Flash (HBF) stack integrated in the GPU package. Such architectures provide a natural location for context-selection acceleration: the retrieved candidate embeddings are already available near the storage array, allowing the optimization objective to be constructed and solved before the candidate data leave the memory subsystem.

\begin{figure}[t]
    \centering
    \includegraphics[width=0.25\textwidth]{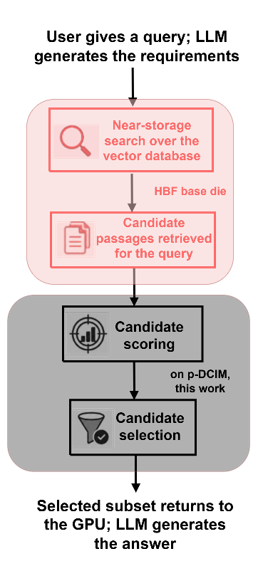}
    \caption{Proposed near-memory RAG context-selection flow. Near-storage retrieval on the HBF stack produces candidate passages, which are scored and jointly selected by the proposed p-DCIM accelerator through QUBO construction and optimization. Only the selected subset is returned to the GPU for LLM answer generation.}
\end{figure}

In this work, we present p-DCIM, a fabricated 28-nm probabilistic SRAM digital compute-in-memory Ising machine for near-memory context selection in RAG. The proposed architecture reuses the same SRAM DCIM array and signed accumulation datapath for two operating modes. In the deterministic mode, candidate embeddings are streamed through the array to evaluate the dot products used to construct the QUBO coefficients. In the probabilistic mode, the resulting coefficients are stored in the same array and the datapath evaluates the local Hamiltonian differences required for sequential Ising updates. Thus, QUBO construction and QUBO optimization share the same storage and arithmetic resources, amortizing access to the candidate embeddings rather than treating coefficient generation and annealing as two independent computational stages.

The probabilistic operation exploits measured pseudo-read errors of the SRAM cells rather than requiring dedicated stochastic devices or external random-number generation. Reducing the SRAM supply voltage $V_{\mathrm{DDM}}$ induces probabilistic disturbances in the stored coefficient bits, and the measured $0\rightarrow1$ and $1\rightarrow0$ error characteristics provide the stochastic perturbation used during annealing. Because the measured voltage dependence contains a narrow transition region that would be difficult to regulate as an analog temperature control in a heterogeneous memory package, we operate the array at a fixed reduced supply and instead control the effective perturbation strength digitally. Successive magnitude bit-planes are protected from pseudo-read disturbance as annealing progresses, while the sign bit remains protected throughout operation. The resulting plane-count schedule provides a discrete cooling mechanism whose stochastic characteristics are directly derived from measurements of the fabricated SRAM macro.

At the system level, we further investigate integration of the proposed accelerator on the custom logic base die of an HBF stack in an HBM/HBF-equipped GPU package. Candidate retrieval is performed near storage, after which the candidate embeddings are consumed locally by p-DCIM for QUBO construction and context selection. Only the query-side information required to construct the objective is supplied by the GPU, and only the selected passage identifiers are returned for final answer generation. Consequently, the $\mathcal{O}(N^2)$ pairwise candidate processing used to construct the set-level objective remains within the memory subsystem rather than traversing the package interconnect. The fabricated macro occupies $0.16~\mathrm{mm}^2$, operates at $500~\mathrm{MHz}$, and consumes approximately $0.41~\mathrm{mW}$ in probabilistic operation. Small QUBO instances are evaluated directly on silicon, while larger HotpotQA instances are evaluated using a measurement-parametrized behavioral model derived from the measured pseudo-read statistics and datapath precision. At the evaluated operating point, context selection requires approximately $2.2~\mathrm{ms}$ and $0.9~\mu\mathrm{J}$, while maintaining answer quality competitive with LLM-based set-selection baselines.

The main contributions of this work are summarized as follows:
\begin{itemize}
    \item We present a fabricated 28-nm dual-mode SRAM-DCIM accelerator that uses a shared memory array and signed accumulation datapath to both construct and solve QUBO objectives for RAG context selection.

    \item We employ a hardware-controlled annealing mechanism based on measured SRAM pseudo-read stochasticity, in which a digitally programmable magnitude-plane exposure schedule provides graded cooling while preserving the coefficient sign.

    \item We map requirement-aware, set-level RAG evidence selection onto the fabricated DCIM architecture and demonstrate the complete execution flow from candidate-embedding scoring and QUBO construction through in-memory Hamiltonian evaluation and passage selection.

    \item We propose a near-memory integration architecture with an HBF-based retrieval system, keeping candidate embeddings and their pairwise interactions near storage and returning only the selected passage identifiers to the GPU.

    \item We characterize the fabricated chip and use a measurement-parametrized model to evaluate larger RAG workloads, demonstrating competitive HotpotQA answer quality while reducing the selection-stage energy by more than six orders of magnitude relative to an LLM-based selector and reducing the evaluated end-to-end query energy by $67\%$.
\end{itemize}

The remainder of this paper is organized as follows. Section II reviews set-level context selection in RAG and the QUBO formulation used in this work. Section III describes the fabricated p-DCIM architecture, including deterministic coefficient construction, probabilistic Hamiltonian evaluation, and the proposed digital annealing schedule. Section IV presents the near-memory system architecture and integration with HBF-based retrieval. Section V reports the measured chip characteristics and evaluates the proposed context-selection flow on RAG workloads. Finally, Section VI concludes the paper and discusses opportunities for scalable near-memory Ising acceleration.

\section{QUBO Formulation} 

Top-$k$ retrieval ranks passages independently and may therefore fail to select a subset that collectively satisfies the query requirements. Following~\cite{singh2026qubooptimizedevidenceselectionretrievalaugmented}, we instead optimize over a subset of the top-$N$ candidates returned by HBF-based near-storage retrieval~\cite{hsu2026havenhighbandwidthflashaugmented}, replacing neural-network-based selection during inference. The Ising hardware considered in this work operates directly in the binary domain and therefore requires the context selection problem to be expressed as a Quadratic Unconstrained Binary Optimization (QUBO) problem. A QUBO is given by
\begin{equation}
    \mathcal{H}_{\mathrm{QUBO}} = \sum_i Q_{ii}q_i + \sum_{i<j}Q_{ij}q_iq_j,
\label{eq:QUBO}
\end{equation}
where $q_i\in\{0,1\}$ are binary decision variables and $\mat{Q}$ contains the corresponding linear and quadratic coefficients. QUBO is equivalent to the Ising formulation through the variable transformation $s_i=2q_i-1$, where $s_i\in\{-1,+1\}$ denotes the corresponding spin variable.

To formulate candidate selection for Retrieval-Augmented Generation (RAG) as a binary optimization problem, each candidate passage $p_i$ is associated with a binary variable $x_i\in\{0,1\}$, where $x_i=1$ indicates selection. Each passage is also assigned a query-relevance score $r_i$. The original query is decomposed by an LLM into a set of requirements, or subqueries, indexed by $j$. The parameter $w_j$ denotes the importance of requirement $j$ with respect to the original query, while $a_{ji}\in\{0,1\}$ indicates whether candidate passage $p_i$ is relevant to requirement $j$. A binary variable $c_j\in\{0,1\}$ indicates whether requirement $j$ is covered by the selected passages. Beyond individual relevance, candidate passages may also interact with one another. Two passages may provide overlapping evidence for the same requirements, making their joint selection redundant, whereas another pair may support different but jointly useful requirements. Since the number of selected passages is limited by the context budget, the selection process should avoid redundant information while favoring complementary evidence. Accordingly, we define $d_{ik}\in[0,1]$ as the pairwise redundancy between passages $p_i$ and $p_k$, accounting for overlap in their information and supported requirements, and $b_{ik}\in[0,1]$ as their pairwise complementarity, capturing the additional utility of selecting passages that support different but jointly useful requirements.

Coverage is enforced by
\begin{equation}
c_j \leq \sum_i a_{ji}x_i,
\label{eq:CoverageConstraint}
\end{equation}
such that $c_j$ can be set to one only if at least one selected passage is relevant to requirement $j$. The context budget is
\begin{equation}
\sum_i x_i \leq K,
\label{eq:ContextConstraint}
\end{equation}
where $K$ denotes the maximum number of passages included in the context. The resulting constrained binary optimization problem is
\begin{equation}
    \begin{aligned}
        \max_{\mathbf{x},\mathbf{c}} \quad
        & \sum_i r_i x_i
        + \sum_j w_j c_j
        - \lambda\sum_i x_i \\
        & - \sum_{i<k} d_{ik}x_i x_k
        + \sum_{i<k} b_{ik}x_i x_k, \\
        \mathrm{s.t.} \quad
        & c_j \leq \sum_i a_{ji}x_i, \qquad \forall j,\\
        & \sum_i x_i \leq K,
    \end{aligned}
\label{eq:ConstrainedObjective}
\end{equation}
The objective balances query relevance and requirement coverage against context size and redundancy, while rewarding complementary evidence. The parameter $\lambda$ controls the cost of including additional passages within the budget $K$. The scores $d_{ik}$ and $b_{ik}$ are normalized to the same range to avoid additional weighting parameters.

The relevance, coverage, and selection-cost terms in~\eqref{eq:ConstrainedObjective} contribute linear coefficients to the QUBO objective, while the redundancy and complementarity terms provide pairwise quadratic coefficients. To remove the remaining constraints, the coverage and context-budget inequalities are converted into equalities using binary-encoded slack variables, and their residuals are added as squared penalties. After negating the original maximization objective and collecting the objective and penalty terms, the context-selection problem is expressed in the standard QUBO form
\begin{equation}
\min_{\mathbf{q}\in\{0,1\}^n}
\mathbf{q}^{\top}\mat{Q}\mathbf{q},
\label{eq:QUBOminimize}
\end{equation}
where $\mathbf{q}$ contains the passage-selection, coverage, and auxiliary slack variables, while $\mat{Q}$ contains the resulting linear and pairwise coefficients. The resulting QUBO is solved by the DCIM datapath described in Section~\ref{sec:architecture}, and the passages corresponding to the optimized selection variables are passed unchanged to the answer generator.

\section{Fabricated p-DCIM Architecture and Dual-Mode Operation}
\label{sec:architecture}

The proposed p-DCIM accelerator is designed to execute the two computational phases required by the context-selection flow of Section II: first, construction of the QUBO coefficients from the retrieved candidate embeddings, and second, iterative solution of the resulting QUBO through stochastic annealing. Rather than implementing these phases using separate similarity-computation and optimization engines, p-DCIM reuses a single SRAM-based DCIM array and its associated signed accumulation datapath across both modes. The deterministic mode computes the embedding-derived quantities used to populate the QUBO matrix, whereas the probabilistic mode stores the resulting quantized coefficients and evaluates the local energy differences required for sequential QUBO updates. This dual-mode reuse is the central architectural feature of the proposed design.

\subsection{Fabricated SRAM-DCIM Macro}

\begin{figure*}[t]
    \centering
    \includegraphics[width=0.75\textwidth]{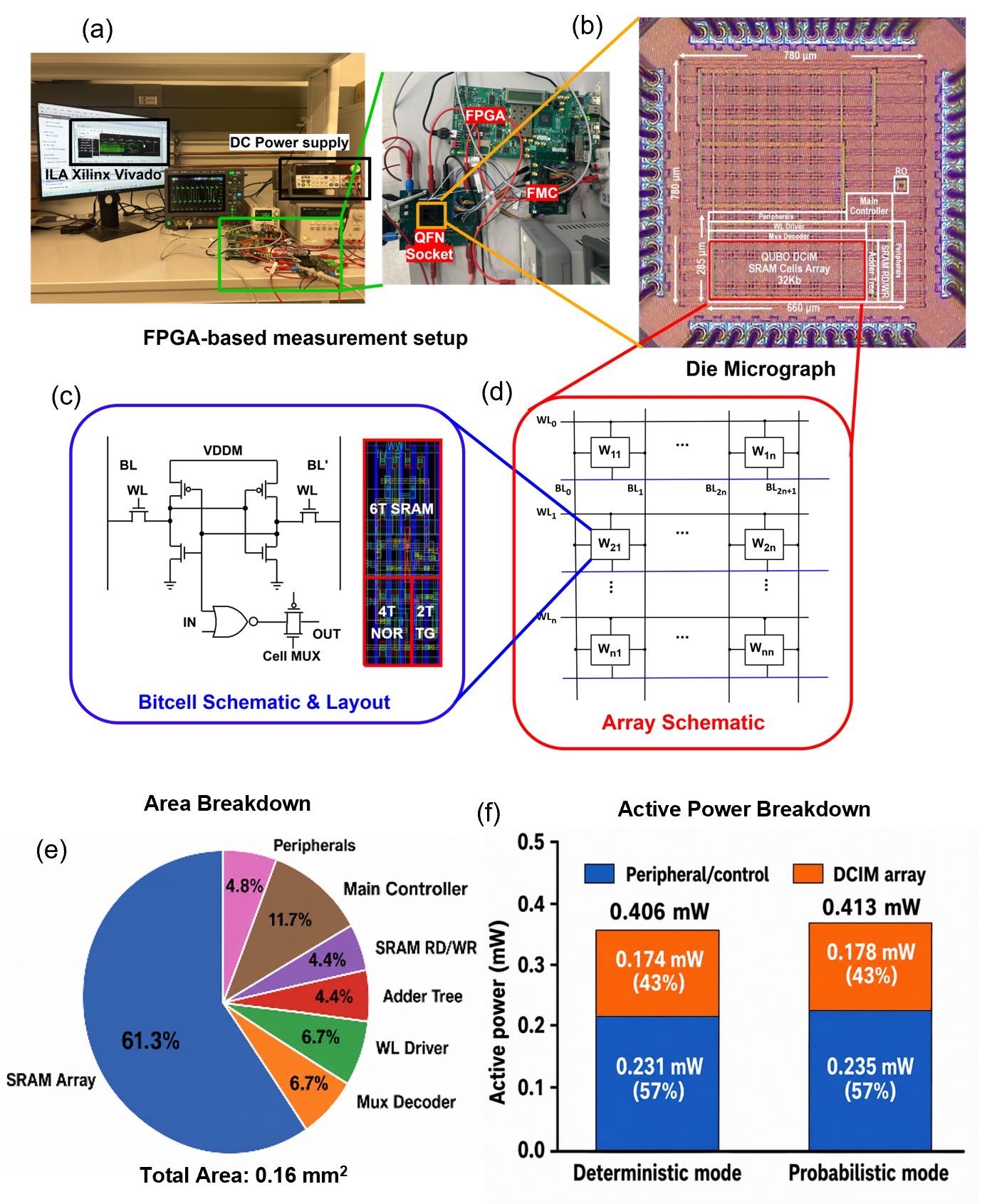}
    \caption{Fabricated 28-nm p-DCIM SRAM Ising accelerator. 
(a) FPGA-based chip measurement setup. 
(b) Die micrograph of the fabricated macro. 
(c) SRAM-DCIM array organization for storing signed, bit-sliced coefficients and performing in-memory accumulation. 
(d) 12T compute cell consisting of a 6T SRAM latch, 4T NOR compute gate, and 2T transmission-gate cell multiplexer. 
(e) Measured area breakdown of the $0.16~\mathrm{mm}^{2}$ macro. 
(f) Measured active-power breakdown in deterministic and probabilistic operating modes, with total powers of $0.406~\mathrm{mW}$ and $0.413~\mathrm{mW}$, respectively.}
\label{fig:chip}
\end{figure*}

Fig.~\ref{fig:chip} shows the fabricated p-DCIM macro and its measurement setup. The chip was fabricated in a 28-nm CMOS process and contains a $64\times64$ logical DCIM array supporting signed 8-bit coefficients. The eight coefficient bit planes are distributed across 512 physical wordlines, resulting in a total SRAM capacity of 32~Kb. The macro occupies $0.16~\mathrm{mm}^{2}$ and operates at a clock frequency of 500~MHz. The fabricated chip is characterized through an FPGA/AXI-based measurement platform, which provides coefficient loading, spin-state control, mode configuration, and readout of the resulting states.

At the cell level, the compute-enabled memory element is a 12T SRAM cell consisting of a conventional 6T storage latch, a 4T NOR compute gate, and a 2T transmission-gate cell multiplexer, as illustrated in Fig.~\ref{fig:chip}(d). The SRAM latch stores one coefficient bit, while the compute gate conditionally enables its contribution according to the applied input. The cell multiplexer selects the active compute path and permits the bit-sliced coefficient planes to be accessed independently. Across a selected row or column, the cell-level outputs are reduced by a signed digital adder tree to form the weighted accumulation required by both operating modes.

The coefficient precision is represented through one protected sign plane and multiple magnitude planes. Conceptually, a signed coefficient can be written as
\begin{equation}
    \widetilde{Q}_{ij}
    =
    s_{ij}
    \sum_{b=0}^{B-2}2^b m_{ij}^{(b)},
    \qquad
    s_{ij}\in\{-1,+1\},
\label{eq:bit_sliced_weight}
\end{equation}
where $m_{ij}^{(b)}\in\{0,1\}$ denotes the stored magnitude bit at plane $b$ and $B=8$ in the fabricated macro. The bit-sliced organization allows the same physical memory to retain the optimization coefficients and participate directly in their arithmetic evaluation.

As shown in Fig.~\ref{fig:chip}(e), the SRAM array dominates the measured macro area, while the remaining area is occupied by the main controller, read/write circuitry, wordline drivers, multiplexing logic, and the signed adder tree. The measured active power is similar in the two operating modes: $0.406~\mathrm{mW}$ in deterministic mode and $0.413~\mathrm{mW}$ in probabilistic mode, as shown in Fig.~\ref{fig:chip}(f). The small difference between the two modes follows from their use of the same physical array and accumulation datapath; probabilistic operation additionally invokes the pseudo-read disturbance mechanism described in Section~\ref{sec:digital_cooling}.

The use of SRAM is particularly suitable for the proposed execution flow. QUBO construction produces a new coefficient matrix for each query, requiring the array to be reprogrammed before annealing. During probabilistic operation, the nominal coefficients are also periodically restored after noise injection. Consequently, the architecture performs frequent writes rather than treating the weight matrix as a stationary parameter set. SRAM provides low write latency and effectively unrestricted write endurance for this operating pattern, while simultaneously supporting the controlled read-disturb mechanism used to generate annealing stochasticity.

\subsection{Deterministic Mode: QUBO Construction}
\label{sec:deterministic_mode}

\begin{figure*}[t]
    \centering
    \includegraphics[width=0.95\textwidth]{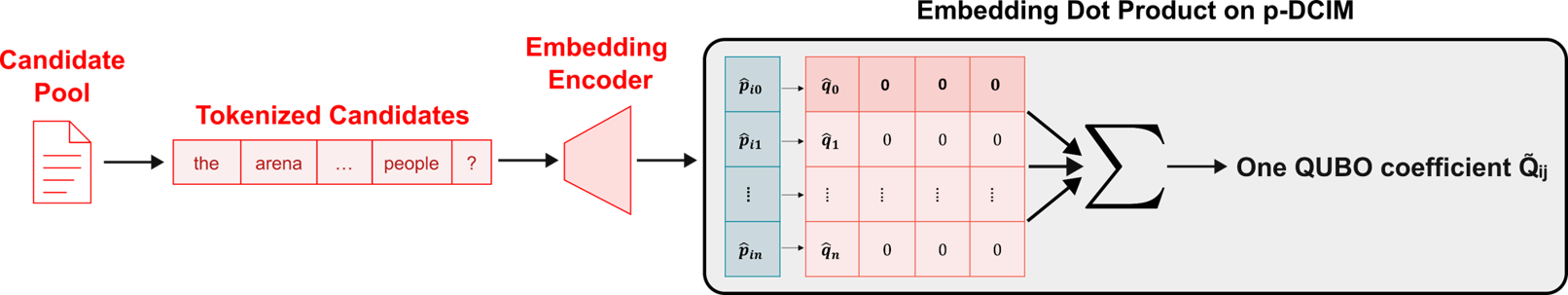}
\caption{Deterministic QUBO-construction mode of p-DCIM. Candidate
passages are tokenized and encoded into embeddings, which are streamed
through the DCIM MAC datapath together with the corresponding query- or
requirement-dependent embedding. The signed accumulation computes the
query--candidate, requirement--candidate, and candidate--candidate
similarity quantities used to form the data-dependent linear and pairwise
QUBO coefficients. Deterministic constraint- and penalty-derived terms
are combined with these coefficients in the host control path before
quantization and write-back into the SRAM array. If the embedding
dimension exceeds the physical array width, partial MAC results are
accumulated across successive passes.}
\label{fig:deterministic_mode}
\end{figure*}

In deterministic mode, the p-DCIM datapath is used to construct the QUBO coefficients required for context selection. As illustrated in Fig.~\ref{fig:deterministic_mode}, candidate embeddings are streamed through the SRAM-DCIM array with the corresponding query or requirement embedding, and the signed adder tree evaluates the dot products used to construct the QUBO coefficients. In this mode, the complete signed MAC output is read from the datapath rather than only its sign.

Let $\mathbf{e}_q$, $\mathbf{e}_i$, and $\mathbf{e}_{u_j}$ denote the normalized embeddings of the query, candidate passage $p_i$, and requirement $j$, respectively. The signed MAC datapath evaluates the dot products required to construct the relevance and requirement-support terms as
\begin{equation}
    \begin{aligned}
        r_i &= \mathbf{e}_q^\top\mathbf{e}_i, \qquad
        w_j = \max\left(0,\mathbf{e}_q^\top\mathbf{e}_{u_j}\right),\\
        a_{ji} &= \mathbb{I}\!\left[\mathbf{e}_{u_j}^\top\mathbf{e}_i \geq \tau_a\right],
    \end{aligned}
\label{eq:deterministic_relevance}
\end{equation}
where $\tau_a$ is the requirement-support threshold. Pairwise passage similarity is obtained as $\sigma_{ik}=\mathbf{e}_i^\top\mathbf{e}_k$ and combined with requirement overlap to form the redundancy coefficient
\begin{equation}
d_{ik} = \max(0,\sigma_{ik}) \frac{\sum_j a_{ji}a_{jk}}{\sum_j a_{ji}+\sum_j a_{jk}-\sum_j a_{ji}a_{jk}}.
\label{eq:deterministic_redundancy}
\end{equation}
The complementarity coefficient $b_{ik}$ is computed from the
requirement-support information following the definition in~\cite{singh2026qubooptimizedevidenceselectionretrievalaugmented},
rewarding passage pairs that contribute evidence toward different but
complementary requirements. Thus, the same MAC datapath provides the similarity quantities used to construct both the linear and pairwise QUBO coefficients.

The remaining constraint-derived coefficients associated with the
slack-variable penalties in Section~II are generated digitally from the
known constraint parameters and combined with the embedding-derived
coefficients before quantization and write-back. Thus, p-DCIM performs
the data-dependent similarity computations that dominate QUBO construction,
while the deterministic algebraic penalty terms are incorporated through
the accompanying host control logic.

\begin{figure*}[t]
    \centering
    \includegraphics[width=0.95\textwidth]{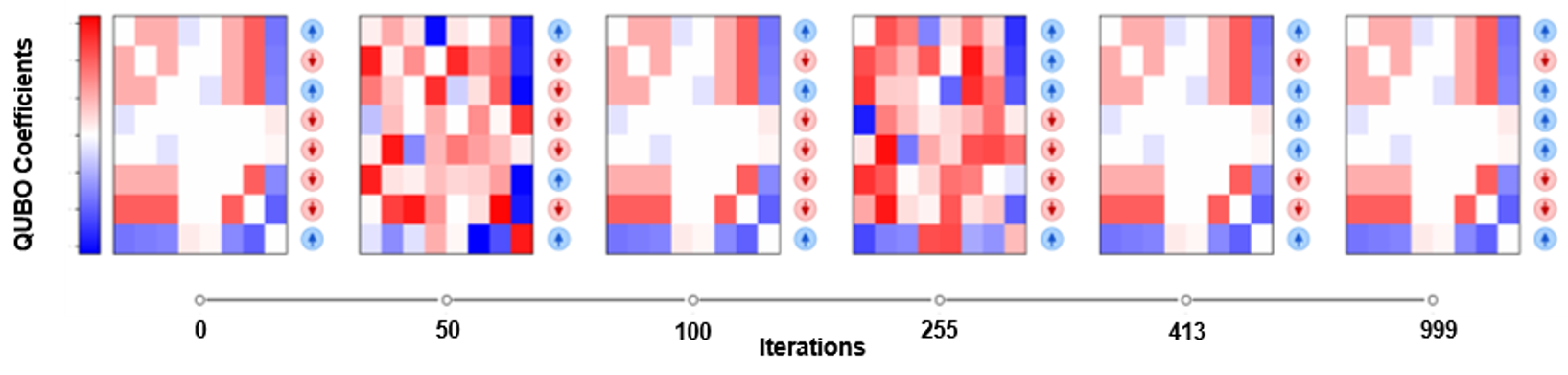}
\caption{Measured evolution of the stored QUBO coefficients during probabilistic operation for a representative 8-variable instance. Pseudo-read operations at reduced $V_{\mathrm{DDM}}$ probabilistically perturb the exposed magnitude bits, providing the stochasticity required for annealing. The nominal coefficients are periodically restored from the on-chip buffer, as illustrated by the refresh at iteration 100, while the optimization state is retained. As the digital plane-count schedule progresses, fewer magnitude planes are exposed to pseudo-read disturbance and the injected coefficient noise decreases toward the end of the anneal.}
\label{fig:probabilistic_mode}
\end{figure*}

\begin{figure*}[t]
    \centering
    \includegraphics[width=1\textwidth]{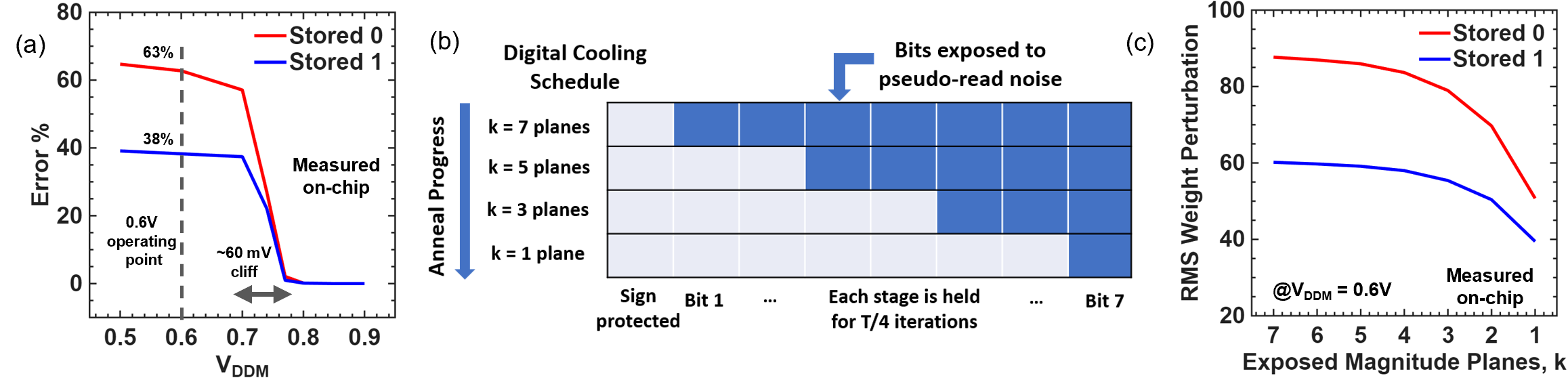}
\caption{Measured SRAM pseudo-read stochasticity and digital annealing control. 
(a) Measured pseudo-read error probability versus $V_{\mathrm{DDM}}$ for cells storing ``0'' and ``1'', showing the asymmetric disturbance probabilities and the approximately $60~\mathrm{mV}$ transition from the high-error regime to near-deterministic operation. The dashed line marks the selected $0.6~\mathrm{V}$ operating point. 
(b) Digital cooling schedule in which the sign plane remains protected and the number of magnitude planes exposed to pseudo-read disturbance is progressively reduced from $k=7$ to $k=1$. 
(c) Measured RMS coefficient perturbation at $V_{\mathrm{DDM}}=0.6~\mathrm{V}$ versus exposed-plane count, demonstrating graded reduction of the effective annealing noise as additional magnitude planes are protected.}
\label{fig:digital_cooling}
\end{figure*}

If the embedding dimension exceeds the physical array width, the dot product is evaluated over multiple MAC passes and the partial results are accumulated. The resulting QUBO coefficients are buffered, quantized, and written back into the same SRAM array for probabilistic operation.

Deterministic operation is performed at $V_{\mathrm{DDM}}=0.9~\mathrm{V}$, where the SRAM operates reliably and pseudo-read-induced disturbances are suppressed. After QUBO construction, the supply voltage is reduced for probabilistic operation, reusing the same SRAM array and signed accumulation datapath for annealing. This eliminates the need for a separate QUBO-construction engine.

\subsection{Probabilistic Mode: In-Memory QUBO Annealing}
\label{sec:probabilistic_mode}

After the QUBO coefficients are constructed and written into the SRAM array, p-DCIM transitions to probabilistic mode to perform the annealing search. The same DCIM array and signed adder tree used for deterministic coefficient construction are now reused to evaluate the local energy change associated with updating each binary variable. For a QUBO objective
\begin{equation}
    E(\mathbf{x})=\mathbf{x}^{T}Q\mathbf{x},
    \qquad x_i\in\{0,1\},
\end{equation}
the update decision for variable $x_i$ is determined by the corresponding local energy difference $\Delta E_i$. The current binary state is applied to the array, the stored QUBO coefficients participate in the in-memory accumulation, and the sign of the resulting $\Delta E_i$ determines whether the visited variable is flipped. Unlike deterministic mode, where the complete signed MAC result is used to form a QUBO coefficient, probabilistic mode requires only the sign of the local energy difference to drive the state update.

For general QUBOs with nonzero diagonal coefficients, we employ the
pinned-one embedding developed in our prior work~\cite{vadlamani2026scalabledigitalcomputeinmemoryising},
which maps each diagonal term to a coupling with an auxiliary variable
clamped to one, allowing the same DCIM MAC datapath to evaluate the
single-variable energy change without a separate diagonal-add path.

Stochasticity is introduced directly into the stored QUBO coefficients through SRAM pseudo-read disturbances at reduced $V_{\mathrm{DDM}}$, rather than through a separate random-number-generation circuit. As illustrated in Fig.~\ref{fig:probabilistic_mode}, the programmed coefficient matrix evolves during annealing as pseudo-read events probabilistically perturb the exposed magnitude bits. The nominal coefficients are periodically restored from an on-chip buffer to prevent uncontrolled accumulation of read-induced errors, while the binary optimization state is retained across the refresh. The example in Fig.~\ref{fig:probabilistic_mode} shows this behavior for a representative 8-variable instance: the coefficients become progressively perturbed during the early iterations, are restored at the programmed refresh point, and subsequently experience decreasing perturbation as the digital plane-count schedule cools the anneal.

Variables are updated sequentially, with each accepted update committed before the next local energy difference is evaluated. The pseudo-read perturbations therefore modify the effective local fields encountered during the scan and provide the stochastic exploration required for annealing without dedicated peripherals such as linear-feedback shift registers (LFSRs). As the number of magnitude planes exposed to pseudo-read noise is progressively reduced, the coefficient perturbations decrease and the system approaches increasingly deterministic operation. The measured pseudo-read characteristics and the corresponding digital cooling mechanism are described in detail in Section~\ref{sec:digital_cooling}.

Thus, the transition from deterministic to probabilistic operation requires no separate optimization datapath: the coefficients constructed using the full signed MAC output are written back into the same array, after which the same array and adder tree are reused for repeated local-energy evaluation during annealing.

\subsection{SRAM Pseudo-Read Stochasticity and Digital Cooling}
\label{sec:digital_cooling}

The stochasticity required for annealing is generated from the intrinsic pseudo-read disturbance of the SRAM cells at reduced supply voltage. Fig.~\ref{fig:digital_cooling}(a) shows the measured pseudo-read error probability as a function of $V_{\mathrm{DDM}}$ for cells initially storing ``0'' and ``1''. The disturbance is asymmetric between the two stored states; at the selected $V_{\mathrm{DDM}}=0.6~\mathrm{V}$ operating point, the measured error probabilities are approximately $63\%$ and $38\%$, respectively. As $V_{\mathrm{DDM}}$ is increased, the error probability falls rapidly toward zero over a narrow transition region of approximately $60~\mathrm{mV}$.

Although this voltage dependence could in principle be used directly as an annealing schedule, the narrow transition makes fine-grained noise control through $V_{\mathrm{DDM}}$ difficult, particularly in the proposed HBM/HBF package environment. We therefore operate the SRAM at a fixed $V_{\mathrm{DDM}}=0.6~\mathrm{V}$, where the pseudo-read disturbance is relatively stable, and control the effective annealing strength digitally through the number of coefficient magnitude planes exposed to pseudo-read noise.

Fig.~\ref{fig:digital_cooling}(b) illustrates the resulting digital cooling schedule. The sign plane is protected throughout annealing so that pseudo-read errors perturb only the coefficient magnitude and do not reverse the polarity of the corresponding interaction. The number of exposed magnitude planes is then progressively reduced as annealing proceeds. In the example schedule, the exposed-plane count is stepped through $k = 7,\;5,\;3,\;1,$ with each stage held for one quarter of the total annealing duration. Early iterations therefore experience stronger coefficient perturbations and greater stochastic exploration, whereas later iterations expose fewer magnitude planes and approach increasingly deterministic operation.

Because the coefficient representation is binary weighted, disturbances in the more significant planes contribute more strongly to the numerical coefficient error. Fig.~\ref{fig:digital_cooling}(c) shows the measured RMS coefficient perturbation at $V_{\mathrm{DDM}}=0.6~\mathrm{V}$ as a function of the number of exposed magnitude planes. The perturbation decreases monotonically as planes are protected, providing a graded effective-temperature schedule without requiring fine analog regulation of the SRAM supply. In particular, removing higher-significance planes produces a substantial reduction in the coefficient disturbance because adjacent bit planes differ by a factor of two in numerical weight.

\begin{figure}[!t]
    \centering
    \includegraphics[width=0.44\textwidth]{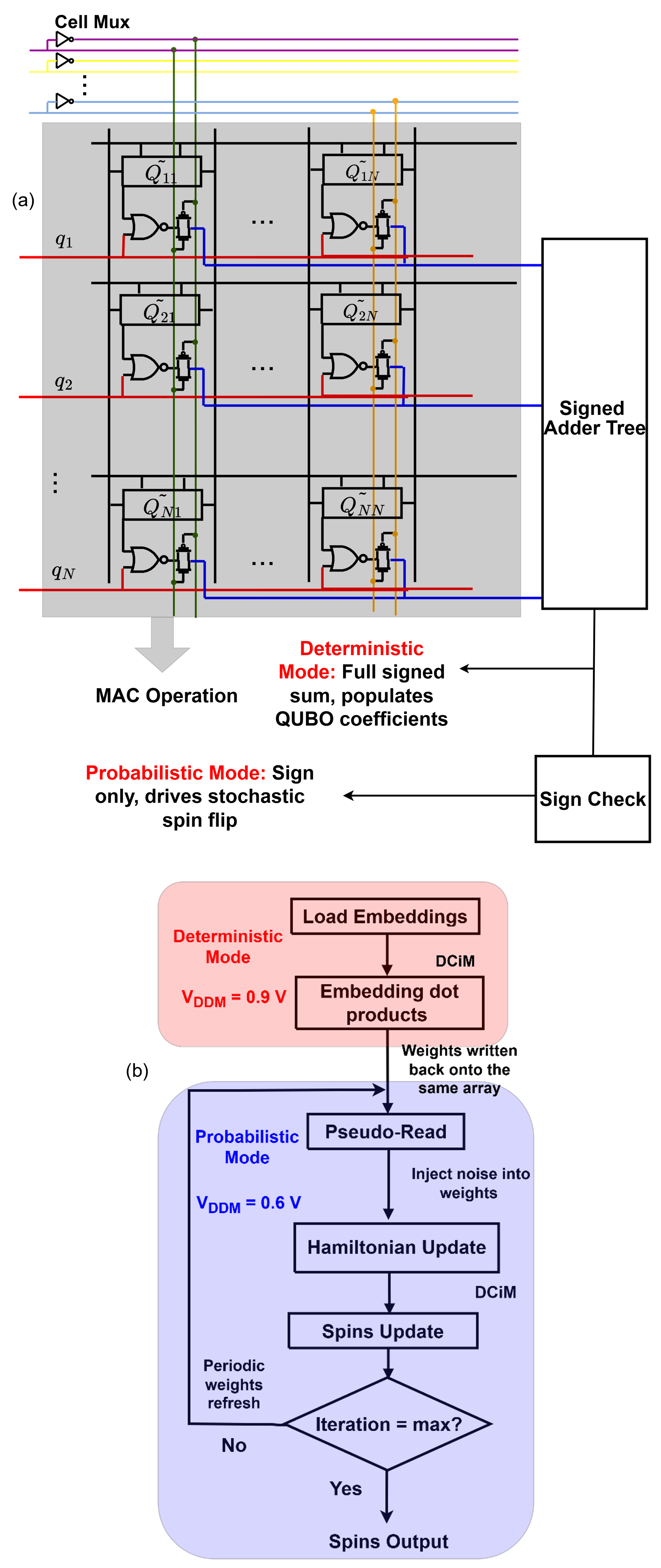}
\caption{Dual-mode operation and control flow of the proposed p-DCIM accelerator. 
(a) Shared SRAM-DCIM array and signed adder tree used in both operating modes. In deterministic mode, the full signed MAC output is used to construct and populate the QUBO coefficients, whereas in probabilistic mode only the sign of the local energy difference is required to determine the spin update. 
(b) End-to-end control flow. The accelerator first operates in deterministic mode at $V_{\mathrm{DDM}}=0.9~\mathrm{V}$ to load embeddings, compute embedding dot products, and write the resulting QUBO coefficients into the SRAM array. The supply is then reduced to $0.6~\mathrm{V}$ for probabilistic operation, where pseudo-read disturbance injects stochasticity into the stored coefficients, followed by in-memory Hamiltonian evaluation and sequential spin updates. The nominal coefficients are periodically refreshed during annealing until the prescribed iteration count is reached.}
\label{fig:dual_mode}
\end{figure}

The proposed scheme therefore separates the source of stochasticity from its annealing control: device-level pseudo-read errors provide the randomness, while simple digital plane-enable signals set its effective strength. This avoids both a finely controlled analog $V_{\mathrm{DDM}}$ sweep and dedicated random-noise-generation circuitry, while retaining a programmable cooling trajectory for the in-memory annealing process.

\subsection{End-to-End Dual-Mode Control Flow}
\label{sec:dual_control}

Fig.~6(b) summarizes the complete p-DCIM execution flow. The
accelerator first operates deterministically at $V_{\mathrm{DDM}}=0.9$ V
to compute, quantize, and program the QUBO coefficients. The supply is
then reduced to $0.6$ V for probabilistic operation, where pseudo-read
disturbance, in-memory local-field evaluation, sequential spin updates,
and periodic coefficient refresh are repeated according to the digital
cooling schedule until the prescribed iteration count is reached.
The final passage-selection state is then returned for downstream
answer generation. Thus, embedding scoring, QUBO solution, and
annealing stochasticity are supported by the same SRAM-DCIM substrate.

\section{Near-Memory RAG System Architecture}
\label{sec:system}

The circuit architecture described in Section~\ref{sec:architecture} enables QUBO construction and optimization to be performed using the same SRAM-DCIM datapath. To exploit this capability at the system level, the accelerator should be placed close to the retrieved candidate embeddings rather than treated as a discrete coprocessor attached to the GPU. In this work, we therefore consider a near-memory integration in which the fabricated p-DCIM architecture is placed on the custom logic base die of a High-Bandwidth Flash (HBF) stack within an HBM/HBF-equipped GPU package. Fig.~\ref{fig:system_arch} contrasts this proposed organization with a conventional RAG execution flow.

\subsection{Data Movement in Conventional RAG Systems}
\label{sec:conventional_rag}

In a conventional RAG system, the vector database resides outside the GPU package, for example in host memory or SSD storage. For each query, candidate passages or their embeddings are retrieved from the external database and transferred across the host interface to the GPU, as illustrated in Fig.~\ref{fig:system_arch}(a). The GPU then performs the subsequent context-selection or reranking operation before invoking the same LLM again for answer generation.

This organization introduces two sources of data movement. First, the retrieved candidate representations must cross the package boundary before the selection decision can be made. Second, when context selection is implemented using an LLM-based set selector, the model weights must be streamed from HBM to the GPU for an additional transformer-inference pass. The result of this relatively expensive computation is nevertheless only a small set of passage identifiers or selected passages that are subsequently supplied to the answer-generation stage.

The cost becomes more pronounced for set-level context selection. Unlike independent relevance ranking, the QUBO formulation considered in this work includes pairwise interactions among candidate passages. For a candidate pool of size $N$, constructing these terms can require $\mathcal{O}(N^2)$ candidate-pair similarity evaluations. Performing this processing after moving the retrieved candidates to the GPU would therefore incur substantial data movement before the Ising optimization itself begins. Consequently, merely attaching an Ising solver to a conventional GPU pipeline does not eliminate the communication associated with constructing the optimization instance.

\subsection{Proposed HBF-Integrated p-DCIM Architecture}
\label{sec:hbf_integration}

\begin{figure*}[t]
    \centering
    \includegraphics[width=0.9\textwidth]{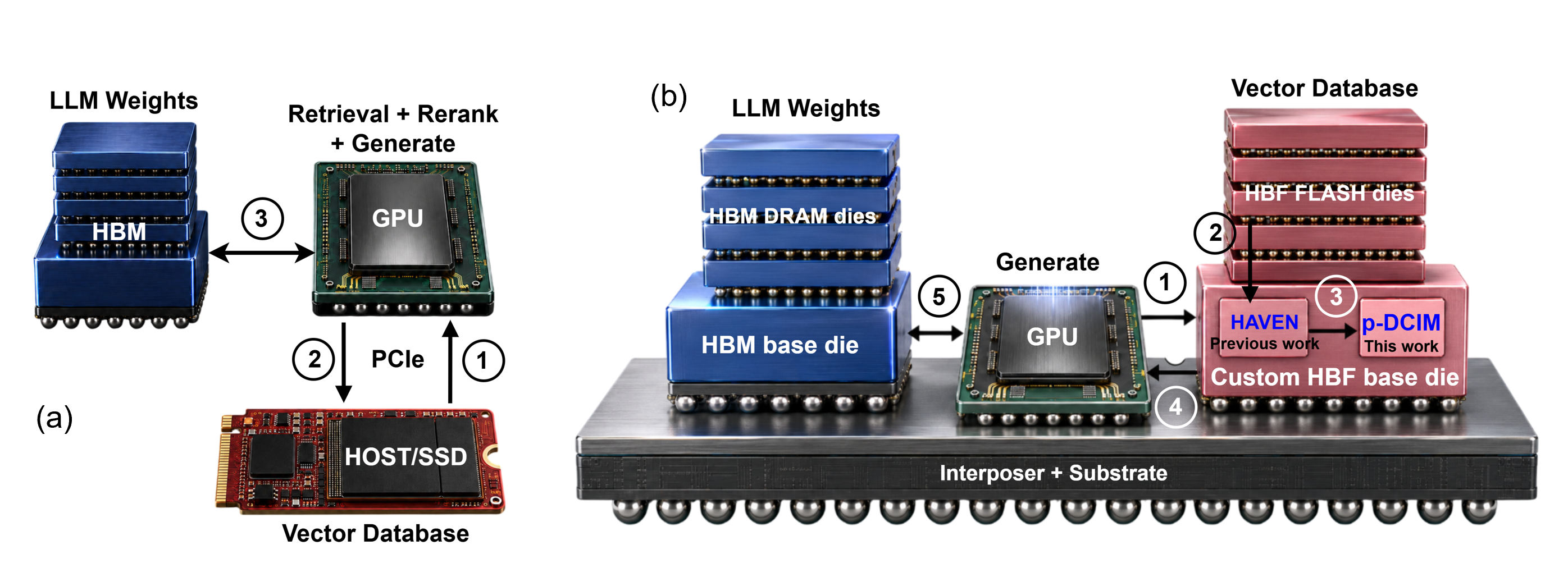}
\caption{Conventional and proposed hardware organizations for RAG context selection.
(a) Conventional RAG execution, where candidate embeddings retrieved from an off-package vector database are transferred to the GPU and context selection is performed using GPU computation and HBM-resident model weights.
(b) Proposed near-memory organization in which the vector database and near-storage retriever reside in an HBF stack and the p-DCIM accelerator is placed on the custom HBF base die. Retrieved candidate embeddings remain within the HBF subsystem while p-DCIM constructs and solves the context-selection QUBO; only the selected passage identifiers are returned to the GPU for final LLM answer generation. The HBM stack continues to store the LLM weights and is not accessed by p-DCIM during context selection.}
\label{fig:system_arch}
\end{figure*}

The proposed system places p-DCIM adjacent to the vector database rather than treating it as a discrete accelerator attached to the GPU. As illustrated in Fig.~\ref{fig:system_arch}(b), we consider an HBF-based near-storage retrieval architecture following HAVEN~\cite{hsu2026havenhighbandwidthflashaugmented}, in which the vector database is stored in the HBF flash dies and candidate retrieval is performed using logic integrated on the custom HBF base die. The p-DCIM accelerator is placed on the same base die and consumes the retrieved candidate embeddings locally. Meanwhile, the HBM stack continues to store the LLM weights, and the GPU performs the semantic operations that require full transformer inference, including query-dependent requirement generation and final answer generation.

This placement is important because the context-selection QUBO contains both query-to-candidate and candidate-to-candidate terms. For a candidate pool of size $N$, the pairwise component can require $\mathcal{O}(N^2)$ similarity evaluations. If the retrieved candidate embeddings were first transferred to the GPU, this pairwise processing would incur substantial package-level data movement before optimization even begins. In the proposed architecture, the retrieved embeddings remain within the HBF subsystem and are consumed directly by p-DCIM. The deterministic mode described in Section~\ref{sec:deterministic_mode} computes the similarity quantities required to form the QUBO coefficients, after which the coefficients are quantized and written back into the same SRAM-DCIM array.

The resulting QUBO is then solved locally using the probabilistic mode of Section~\ref{sec:probabilistic_mode}. Because QUBO construction and optimization reuse the same physical array and adder-tree datapath, neither the candidate embeddings nor the complete QUBO matrix need to be transferred to a separate accelerator during context selection. Only the final passage-selection result is returned to the GPU for downstream generation.

The end-to-end query flow therefore proceeds as follows. First, the GPU generates the query-dependent information required for retrieval and context selection and sends it to the HBF subsystem. Second, the near-storage retrieval engine searches the HBF-resident vector database and produces the candidate pool locally. Third, p-DCIM constructs the QUBO from the retrieved candidate embeddings and solves it using the dual-mode execution described in Section~\ref{sec:architecture}. Fourth, the selected passage identifiers are returned to the GPU. These identifiers are subsequently used to fetch only the selected
$K$ passage contents required for generation; the remaining candidate
passages and embeddings remain within the HBF subsystem. Finally, the GPU assembles the selected context and performs answer generation using the LLM weights stored in HBM.

This organization keeps the high-dimensional candidate representations and their pairwise processing close to storage while reserving the GPU for full LLM inference. The $\mathcal{O}(N^2)$ candidate-interaction workload remains on the HBF base die, and only a compact selection result crosses back to the GPU after optimization. Fig.~\ref{fig:system_arch}(b) represents a proposed package-level integration of the fabricated p-DCIM macro; the HBM/HBF-integrated system itself is not fabricated in this work. The resulting latency and energy implications are evaluated in Section~\ref{sec:results}.

\section{Experimental Results and System Evaluation}
\label{sec:results}

We evaluate the proposed system at both the circuit and application levels. Small QUBO instances that fit within the fabricated $64\times64$ p-DCIM macro are executed directly on chip, while larger RAG context-selection instances are evaluated using a measurement-parametrized behavioral model derived from the measured SRAM pseudo-read statistics and the finite-precision datapath of the fabricated chip. At the application level, we evaluate whether the selected contexts preserve downstream answer quality on HotpotQA, and at the system level we compare the latency and energy of p-DCIM context selection against an LLM-based set-selection baseline executed on a GPU.

\subsection{Experimental Methodology and End-to-End Demonstration}
\label{sec:results_setup}

The fabricated p-DCIM macro is characterized using the FPGA/AXI measurement setup described in Section~\ref{sec:architecture}. QUBO instances that fit within the physical array are executed directly on chip, while larger RAG context-selection instances are evaluated using a measurement-parametrized behavioral model incorporating the measured SRAM pseudo-read statistics, signed 8-bit coefficient quantization, sequential spin updates, periodic coefficient refresh, and the digital plane-count cooling schedule of Section~\ref{sec:digital_cooling}. This enables application-scale evaluation while retaining the measured stochastic and finite-precision characteristics of the fabricated hardware.

\begin{figure*}[t]
    \centering
    \includegraphics[width=0.9\textwidth]{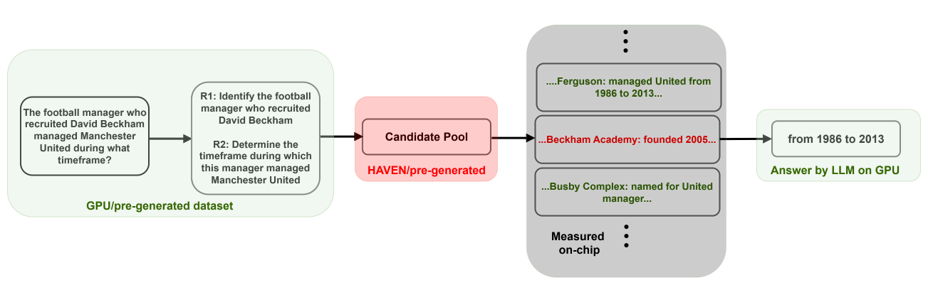}
\caption{Representative end-to-end HotpotQA context-selection example. 
The query and information requirements are generated on the GPU, and a pre-generated candidate pool is supplied to the p-DCIM selection stage. The fabricated chip performs the context-selection QUBO on-chip, retaining passages that jointly provide the required evidence (green) while rejecting an irrelevant candidate (red). The selected context is returned to the GPU, where the LLM generates the final answer.}
\label{fig:rag_example}
\end{figure*}

For the RAG evaluation, the query, information requirements, and candidate passages are generated before the p-DCIM selection stage; requirement generation and final answer generation remain on the GPU. Fig.~\ref{fig:rag_example} illustrates a representative HotpotQA example. The fabricated chip selects complementary passages required to answer the query while rejecting an irrelevant candidate, after which the selected context is returned to the GPU for answer generation. This example demonstrates the intended role of p-DCIM as the intermediate combinatorial context selector in an otherwise conventional RAG pipeline.

\begin{figure*}[t]
    \centering
    \includegraphics[width=1\textwidth]{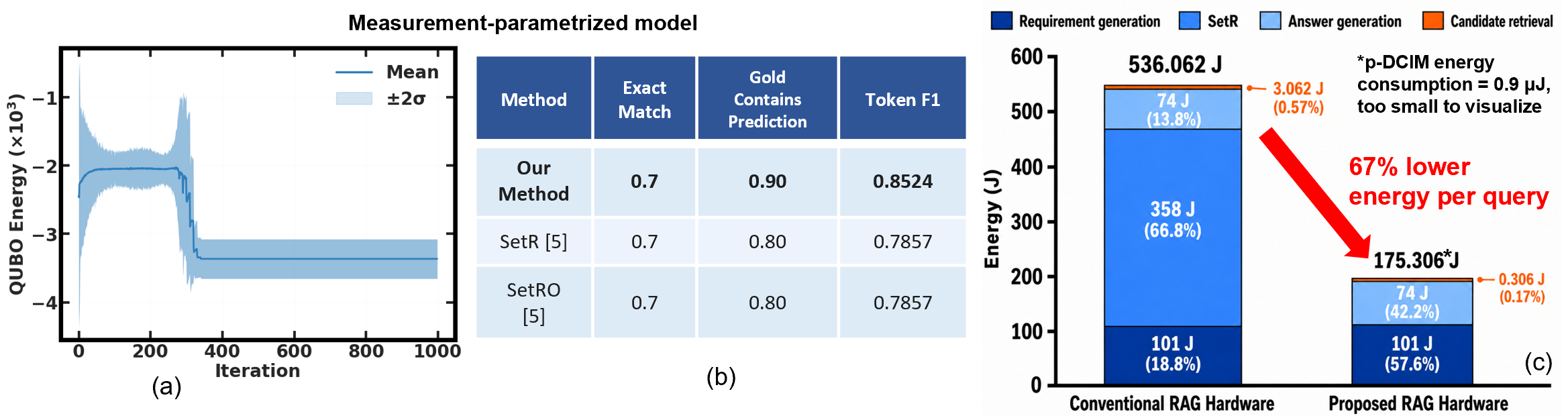}
\caption{Application and system-level evaluation of the proposed p-DCIM context-selection flow.
(a) QUBO energy versus annealing iteration for a representative HotpotQA instance, showing the mean and $\pm2\sigma$ variation across repeated runs of the measurement-parametrized model derived from measured chip characteristics.
(b) Downstream answer quality compared with SetR and SetRO \cite{lee-etal-2025-shifting} using Exact Match, Gold-Contains, and token-level F1 metrics.
(c) Energy per query by RAG pipeline stage. Requirement generation and answer generation remain on the GPU in both systems, while the LLM-based selection pass is replaced by p-DCIM, reducing selection energy to approximately $0.9~\mu\mathrm{J}$ and the evaluated total query energy by $67\%$.}
\label{fig:results}
\end{figure*}

\subsection{QUBO Solution and RAG Answer Quality}
\label{sec:qubo_quality}

The fabricated macro executes the complete annealing loop for QUBO
instances that fit within the physical array; across all ten measured
instances, the minimum-energy state reached by p-DCIM matched the
brute-force optimum.
Larger HotpotQA-derived QUBOs are evaluated using the measurement-parametrized model. As shown in Fig.~\ref{fig:results}(a), the mean QUBO energy decreases
during annealing and stabilizes as the digital plane-count schedule reduces
the injected coefficient perturbation.

The resulting passage selections are evaluated through downstream answer
generation. Fig.~\ref{fig:results}(b) compares the proposed method with
SetR and SetRO~\cite{lee-etal-2025-shifting}. On the evaluated HotpotQA examples,
p-DCIM achieves an Exact Match score of $0.70$, Gold-Contains of $0.90$,
and token-level F1 of $0.8524$, compared with $0.70$, $0.80$, and $0.7857$,
respectively, for both LLM-based selectors. These results demonstrate that
the QUBO-based selector can replace the LLM set-selection pass while
preserving competitive downstream answer quality. The present evaluation
is intended as a hardware-system demonstration; a larger-scale evaluation
of the underlying QUBO selection formulation is reported separately
in~\cite{singh2026qubooptimizedevidenceselectionretrievalaugmented}.

\subsection{Selection Latency and Energy}
\label{sec:energy_results}

For the evaluated workload, p-DCIM completes context selection in approximately
$2.2~\mathrm{ms}$. Using the measured probabilistic-mode power of
$0.413~\mathrm{mW}$, this corresponds to approximately $0.9~\mu\mathrm{J}$
per selection. The resulting $0.9~\mu\mathrm{J}$ should therefore be interpreted as a
macro-normalized estimate based on the measured power of the fabricated
array, rather than a full-array power projection. In comparison, the SetR selection stage requires an additional
LLM inference pass on an NVIDIA RTX A6000 and consumes approximately
$358~\mathrm{J}$ per query. As shown in Fig.~\ref{fig:results}(c), replacing
this LLM-based selection pass with p-DCIM reduces the selection-stage energy
by more than six orders of magnitude.

Requirement generation and final answer generation remain unchanged and are
executed on the GPU in both systems. Under the evaluated workload, replacing
only the selection stage reduces the total query energy from approximately
$536~\mathrm{J}$ to $175~\mathrm{J}$, corresponding to a $67\%$ reduction.
The p-DCIM power and stochastic characteristics are measured from the
fabricated chip, while HotpotQA-scale selection and the proposed HBF-integrated
system are evaluated using the measurement-parametrized model described above.

\section{Conclusion}
\label{sec:conclusion}

This work presented p-DCIM, a fabricated 28-nm probabilistic SRAM
digital compute-in-memory Ising accelerator for near-memory context
selection in RAG. The proposed dual-mode architecture reuses the same
SRAM array and signed accumulation datapath to construct and solve the
context-selection QUBO: deterministic operation evaluates
embedding-derived similarity terms, while probabilistic operation performs
in-memory local-energy evaluation for annealing. SRAM pseudo-read
disturbances provide the required stochasticity without dedicated
random-noise-generation peripherals, while a digital magnitude-plane
schedule controls the effective annealing strength. Measurements of the
fabricated macro demonstrate the pseudo-read characteristics, dual-mode
operation, and on-chip solution of small QUBO instances.

At the system level, we proposed placing p-DCIM on the logic base die of
an HBF-based near-storage retrieval system, allowing retrieved candidate
embeddings to remain near storage during QUBO construction and selection.
HotpotQA-scale instances are evaluated using a measurement-parametrized model
derived from the measured chip characteristics. The proposed flow preserves
competitive downstream answer quality while reducing the evaluated
selection-stage energy from an LLM-based GPU selection pass to
approximately $0.9~\mu\mathrm{J}$ and reducing total query energy by
$67\%$. These results demonstrate how a general QUBO-capable DCIM
substrate can extend Ising acceleration from optimization alone to
application-dependent construction and solution of practical AI workloads.

\section*{Acknowledgments}
AI Chatbot tools were used to assist with grammar and figure polishing; all scientific content and analysis were created solely by the authors. This work was supported in part by NSF Grant 2311295.

\bibliographystyle{IEEEtran}
\bibliography{ref.bib}

\end{document}